\documentclass[usenatbib,useAMS,letters]{mnras}

\usepackage{natbib}
\usepackage{amsmath}
\usepackage{graphicx}
\usepackage{color}
\usepackage{mathrsfs}
\usepackage{epsfig}
\usepackage{comment}
\usepackage{ulem}

\usepackage{graphicx}
\usepackage[hang,small,bf]{caption}
\usepackage[subrefformat=parens]{subcaption}
\usepackage{amssymb}
\usepackage[T1]{fontenc}

\newcommand{\msun}{{M}_{\odot}}

\newcommand{\rsun}{{\rm R}_\odot}

\newcommand{\beq}{\begin{equation}}
\newcommand{\eeq}{\end{equation}}

\defcitealias{K14}{K14}
\defcitealias{K16}{K16}
\defcitealias{Visbal_2015}{VHB15}

\usepackage{url} %It provides better support for handling and breaking URLs.

\title[Formation of Binary Black Holes Similar to GW190521
from Population III Binary Star Evolution]
{Formation of Binary Black Holes Similar to GW190521 with a Total Mass of $\sim 150\,M_{\odot}$ from Population III Binary Star Evolution}

\author[T. Kinugawa et al.]
{Tomoya Kinugawa$^{(1)}$\thanks{E-mail: kinugawa@icrr.u-tokyo.ac.jp},  Takashi Nakamura$^{(2)}$, and Hiroyuki Nakano$^{(3)}$\\
\\
$^{1}$Institute for Cosmic Ray Research, The University of
  Tokyo, Kashiwa, Chiba 277-8582, Japan\\
$^{2}$Department of Physics, Graduate School of Science, Kyoto University,
Kyoto 606-8502, Japan\\
$^{3}$Faculty of Law, Ryukoku University, Kyoto 612-8577, Japan}

\begin{document}

\date{\today}
\maketitle

\begin{abstract}
 In the case of zero-metal (population III or Pop III) stars, we show that the total mass of
 binary black holes from binary Pop III star evolution can be $\sim 150 \,M_{\odot}$, which agrees with the mass of the binary black hole GW190521 recently discovered by LIGO/Virgo. {The event rate of such binary black hole mergers is estimated
 as 0.13--0.66$~(\rho_{\rm SFR}/(6\times10^5~\msun/{\rm Mpc}^3))~Err_{\rm sys}~{\rm yr^{-1}~Gpc^{-3}}$,
 where $\rho_{\rm SFR}$ and $Err_{\rm sys}$ are the cumulative comoving mass density of Pop III stars depending on star formation rate and the systematic errors depending on uncertainties in the Pop III binary parameters, respectively. The event rate in our fiducial model with $\rho_{\rm SFR}=6\times10^5~\msun/{\rm Mpc}^3$ and $ Err_{\rm sys}=1$  is 0.13--0.66$~{\rm yr^{-1}~Gpc^{-3}}$, which
  is consistent with the observed value of 0.02--0.43$~{\rm yr^{-1}~Gpc^{-3}}$}.
\end{abstract}


%%%%%%%%%%%%%%%%%%%%%%%%%%%%%%%%%%%%%%%
\section{Introduction}
%%%%%%%%%%%%%%%%%%%%%%%%%%%%%%%%%%%%%%%

GW190521, observed in the LIGO/Virgo third observing run
(O3a)~\citep{Abbott:2020tfl,Abbott:2020mjq}, is a gravitational-wave (GW) signal from
a merging binary black hole (BH)
with a primary BH mass of 71--106$\,M_{\odot}$~\footnote{Here, all values estimated by LIGO/Virgo are shown using the symmetric 90\% credible interval.},
and a secondary BH mass of 48--83$\,M_{\odot}$.
The remnant BH after the merger has a mass 
of 126--170$\,M_{\odot}$,
and thus this object can be considered as
an intermediate-mass BH in the mass range 100--1000$\,M_{\odot}$.
The redshift of GW190521 is 0.48--1.1, while
the merger rate density is estimated as 0.02--0.43$~{\rm yr^{-1}~Gpc^{-3}}$.

Here, we should note that the two component masses of GW190521
are possibly within the pair-instability supernova (PISN) mass gap
for metallicity $Z \geq 0.001\,Z_{\odot}$. In~\cite{Woosley:2016hmi}, this PISN mass gap 
is described as 
``No black holes between 52 and 133$\,M_{\odot}$
are expected from stellar evolution in close binaries''.
In more detail, for example, \cite{Leung:2019fgj} discussed 
pulsational PISNe (PPISNe) for 
$Z \geq 0.001\,Z_{\odot}$ by simulating 
a helium core without the hydrogen envelope,
as simulation of the whole star
is computationally expensive. They obtained a lower bound of the PISN mass gap
 of approximately 50$\,M_{\odot}$, indicating that merging of two BHs with mass
$\lesssim 50\,M_{\odot}$ is needed 
to form a BH with a mass $\gtrsim 50\,M_{\odot}$
\cite[e.g.,][]{Fragione:2020aki}.

However, the upper limit of the mass of BHs for population (Pop) III stars with $Z=0$ is different from that for $Z \geq 0.001\,Z_{\odot}$.
Because Pop III stars do not tend to lose the envelope,
they will have a different lower bound of the PISN mass gap to that of Pop II stars.
The CO core mass ($M_{\rm CO}$) range of PPISNe is approximately 40--60$\,\msun$
\citep{Heger2002,Heger2003,Umeda2008,Waldman2008,Yoshida2016}.
Calculations of Pop III star evolution \citep{Marigo_2001,Heger2002,Heger2003,Ekstrom_2008,Tanikawa2019} show
that to form such a massive CO core, the zero-age main sequence (ZAMS) mass of Pop III stars has to exceed $\sim100\,\msun$.
Furthermore, \cite{Yoon:2012aq} and~\cite{Chatzopoulos:2012fz}
computed the evolution of Pop III metal-free ($Z=0$)
massive stars, and Table 1 in~\cite{Chatzopoulos:2012fz}
shows that a nonrotating Pop III star with a ZAMS mass of 
$M_{\rm ZAMS} = 75\,M_{\odot}$ becomes a BH in the core collapse.
Thus, the formation of BHs with $M \lesssim 80\,M_{\odot}$
 from Pop III stars is a possible natural explanation of the existence of 
GW190521-like binary BHs for binary Pop III stars
(see Figure~12 in~\cite{Yoon:2012aq}
and Figure~5 in~\cite{Chatzopoulos:2012fz}).

We should note that many proposals
and discussions
to explain BHs within the PISN mass gap
and their dynamics were made immediately after the announcement of
GW190521:
{primordial BHs~\citep{Carr:2019kxo,DeLuca:2020sae,Vovchenko:2020crk},
% no estimation of the merger rate,
% R(M1,2 > 65Msun) \eqsim 1.1/yr, R(M1 > 85Msun, M2 > 65Msun) \eqsim 0.8/yr,
% no estimation of the merger rate,
numerical relativity simulations on extremely eccentric (head-on) mergers~\citep{CalderonBustillo:2020odh},
% no estimation of the merger rate
numerical relativity simulations on a high-eccentricity, precessing model~\citep{Gayathri:2020coq},
% no estimation of the merger rate
GW data analysis using an eccentric waveform model~\citep{Romero-Shaw:2020thy},
% no estimation of the merger rate
possible existence beyond the Standard Model explanations~\citep{Sakstein:2020axg},
% no estimation of the merger rate
an analysis of BH mass functions based on the ten binary BHs observed in GWTC-1~\citep{Wang:2020aoh},
% no estimation of the merger rate
BH masses in a modified gravitational theory~\citep{Moffat:2020vdu},
% no estimation of the merger rate
repeated BH mergers in star clusters~\citep{Fragione2020},
% only probability, no estimation of the merger rate
head-on collisions of two horizonless
vector boson stars~\citep{CalderonBustillo:2020srq}, and
% no estimation of the merger rate
binary BHs straddling the PISN mass gap~\citep{Fishbach:2020qag}.
% no estimation of the merger rate
Of the above references, \cite{DeLuca:2020sae}
estimated the event rate of GW190521-like binary BHs
as $\eqsim 1.1~{\rm yr}^{-1}$ for ($M_1$, $M_2 > 65\msun$)
and $\eqsim 0.8~{\rm yr}^{-1}$ for ($M_1 > 85\msun$, $M_2 > 65\msun$) in a primordial BH scenario, assuming the detectability threshold (signal-to-noise ratio (SNR) $= 8$) for the O3 sensitivity.
Here, $M_1$ and $M_2$ are the masses of the primary
and secondary objects, respectively.}
%Also we note that \cite{Farrell2020} have discussed ...}
% Farrell2020: temporally input

{On the other hand, the existence of mass-gap BHs like GW190521 has been suggested by Pop III binary star evolution  \citep{Kinugawa:2015nla,Kinugawa2020,Tanikawa2020},
which is supported by a recent calculation of Pop III stellar evolution \citep{Farrell2020}.
However, the event rate has not yet been discussed, including the various models shown in the previous paragraph, except for a primordial BH model by \cite{DeLuca:2020sae}.
In this Letter, we discuss the formation process of GW190521-like binary BHs from Pop III stars
and estimate the event rate from population synthesis simulations of Pop III
binary stars.}

%%%%%%%%%%%%%%%%%%%%%%%%%%%%%%%%%%%%%%%
\section{Analysis}
%%%%%%%%%%%%%%%%%%%%%%%%%%%%%%%%%%%%%%%

First, we discuss the effect of the rotational velocity
of Pop III stars at ZAMS.
When the end state of a Pop III star is described by a Kerr BH,
the angular momentum of the BH should satisfy the following inequality:
\begin{equation}
J_{\rm BH} < M \frac{GM}{c} \,.
\end{equation}
Assuming no angular momentum loss of the star up to the formation of BH,
we may regard the angular momentum at ZAMS as equal to that of the final BH, that is, $J_{\rm ZAMS} = J_{\rm BH}$. The Pop III star at ZAMS has the rotational energy
\begin{align}
E_{\rm rot} &\sim \frac{J_{\rm ZAMS}^2}{I}
\cr
& = \frac{J_{\rm BH}^2}{I} \,,
\end{align}
where $I$ is the moment of inertia of the Pop III star.
The gravitational energy of the star is written as
\begin{equation}
E_{\rm grav} \sim \frac{GM^2}{R} \,,
\end{equation}
where $R$ is the stellar radius.
By comparing the rotational energy (Eq.~(2)) with the gravitational energy (Eq.~(3))
and using Eq.~(1), we have
\begin{align}
\frac{E_{\rm rot}}{E_{\rm grav}}
& <  \frac{GM^2R}{I c^2} \cr
&= \frac{GM}{c^2R\kappa} \,,
\end{align}
where $\kappa$ is defined by $I=\kappa MR^2$.
Using the values of $M$, $R$, and $\kappa$
for Pop III stars at ZAMS,
we can estimate the upper limit of the effect of rotation.
The stellar radius at ZAMS ($R_{\rm{ZAMS}}$) is given by~\cite{Kinugawa2014} as
\begin{align}
\frac{R_{\rm{ZAMS}}}{{\rm R}_\odot} = &
1.22095+2.70041\times 10^{-2} \left(\frac{M}{10\,\msun}\right)
\cr &
+0.135427\left(\frac{M}{10\,\msun}\right)^2
\cr &
-1.95541\times10^{-2}\left(\frac{M}{10\,\msun}\right)^3
\cr &
+8.7585\times10^{-4}\left(\frac{M}{10\,\msun}\right)^4 \,.
\end{align}
Using Eq.~(5), for $M_{\rm{ZAMS}} = 10$--$100\,\msun$ we have
\begin{equation}
\frac{GM}{c^2R} \sim 
1.6\times 10^{-5} - 5\times 10^{-5} \,.
\end{equation}
From Eqs.~(6) and (4), if $\kappa \gg 5\times 10^{-5}$,
we may ignore the effect of rotation. Thus, we can use the results of 
the evolution of spherically symmetric Pop III stars.
In practice, $\kappa = 0.21$ for the core,
and $O(0.1)$--$O(0.01)$ for the outer layer \citep{Hurley_2002,Kinugawa2014};
hence, the rotational energy is at most $0.01$ times the gravitational energy~\footnote{For polytropes, we have $\kappa=0.4,\,0.261,\,0.155,\,0.0754,\,0.0226$,
and $0.00690$ for the polytropic indexes of
$n=0,\,1,\,2,\,3,\,4$, and $4.5$, respectively.}.
Therefore, we can use the results of 
the evolution of spherically symmetric Pop III stars as a good approximation to rotating ones.

Next, we discuss the evolution of Pop III binaries.
In our previous work~\citep{Kinugawa2020},
we simulated the evolution of Pop III binaries 
using population synthesis simulations for various models
with different initial conditions: 
initial mass function, initial mass ratio,
separation and eccentricity distributions of binaries, 
binary evolution parameters such as 
mass transfer rate of the donor,
fraction of transferred stellar mass in the accretion process,
common envelope parameters,
and tidal coefficient factor.
As a result, we found that the chirp mass~\footnote{Here, the chirp mass is defined by
\begin{equation}
M_{\rm chirp} = \frac{(M_1 M_2)^{3/5}}{(M_1+M_2)^{1/5}} \,.
\end{equation}}
distribution of binary BHs formed from Pop III star binaries has
a peak at $M_{\rm chirp} \sim 30\,M_{\odot}$
and the merger rate density of the Pop III binary BHs
at $z=0$
is in the range 3.34--21.2$~{\rm yr^{-1}~Gpc^{-3}}$ for seven different models.
This is consistent with the LIGO/Virgo result of 
9.7--101$~{\rm yr^{-1}~Gpc^{-3}}$~\citep{LIGOScientific:2018mvr}.

In~\cite{Kinugawa2020}, we considered only the PISNe without  the remnant if the CO core mass is more than 
$60\,M_{\odot}$, corresponding to the initial total mass  $M_{\rm initial}$ of $\sim130\,M_{\odot}$~\citep{Heger2002}.
In this Letter, we also consider the PPISNe when the CO core mass is between 40$\,\msun$ and 60$\,\msun$ \citep[e.g.,][]{Yoshida2016}.

In previous studies on the mass ejection of PPISNe \citep[e.g.][]{Yoshida2016,Leung:2019fgj},
only Pop II stars that had already lost the hydrogen-rich envelope before PPISN were considered.
In contrast, Pop III PPISN progenitors tend to keep 
the hydrogen envelope due to the lack of metals, and thus the envelope mass may be halted.
In this circumstance, we consider two possible extreme models and consider the reality to be between these two extremes.
One is the model without mass ejection at PPISN (no (mass ejection) PPISN model).
This model is the same as the fiducial model of \cite{Kinugawa2020}.
In the other extreme case, the hydrogen and helium envelopes are totally ejected by PPISN, and the remnant mass is equal to the CO core mass (PPISN model). 
We have two extreme estimations of the event rate of GW190521-like binary BH mergers from the above two extreme models: the no-mass ejection model gives the upper bound of the event rate, whereas the PPISN model gives the lower bound.

\begin{figure}
    \centering
    \includegraphics[width=0.47\textwidth]{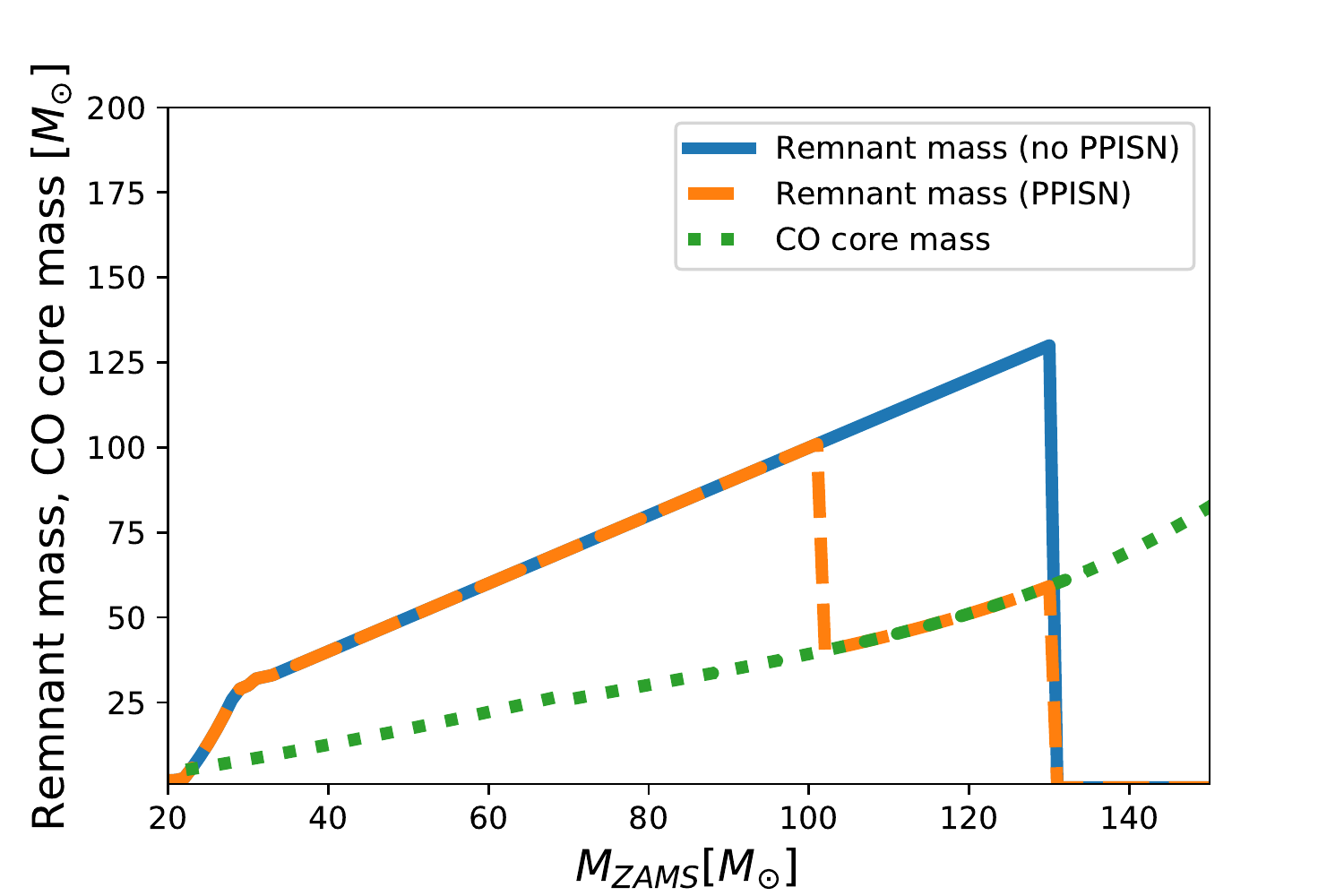}
    \caption{Remnant mass and CO core mass as a function of the zero-age main-sequence mass of the progenitor star.
    The blue solid line shows the model without mass ejection at PPISN (no PPISN), and the orange dashed line shows the model with mass ejection (PPISN).
    The green dotted line is the CO core mass.
    }
    \label{fig:mass}
\end{figure}

Figure~\ref{fig:mass} shows the remnant mass and CO core mass as a function of the ZAMS mass of the progenitor star.
In this figure, the blue solid and orange dashed lines show the models without mass ejection at PPISN (no PPISN) and with mass ejection (PPISN), respectively.
The CO core mass is shown by the green dotted line.
Although binaries can lose the envelope via the binary interaction, a single Pop III star can evolve up to $\sim100\,\msun$ BH even in the worst case, which corresponds to the orange dashed line in Figure~\ref{fig:mass}.  

\begin{table*}
\caption{Distribution functions of initial conditions in evolution of Pop III binaries. Functional forms are shown for the initial mass function ($M_1$),
distribution of initial mass ratio ($q$), distribution of orbital separation ($a$),
and eccentricity ($e$) distributions of binaries
with the range of the parameter.
$M_1$ denotes the primary mass 
and $q=M_1/M_2$ is the mass ratio,
where $M_2$ denotes the secondary mass.}
%For details see \cite{Kinugawa:2020ego}.}
\label{table:initial}
\centering
\begin{tabular}{c c c c}
	\hline
	Initial mass function & Initial mass ratio & Initial separation & Initial eccentricity\\
	\hline
	flat  & flat  & logflat   & Power law (index:1) \\ 
	$10\msun<M_1<150\msun$  & $10\msun/M_1<q<1$ & $\log a_{\rm min}<\log (a/\rsun)<6$  & $0<e<1$ \\
	\hline
\end{tabular}	
\end{table*}

The distribution functions of the initial conditions for evolution of Pop III binaries for our models 
are summarized in Table~\ref{table:initial}
\footnote{See~\cite{Kinugawa2020} for details of the binary evolution.}.
The functional forms are shown for the initial mass function ($M_1$),
distribution of initial mass ratio ($q=M_1/M_2$), distribution of orbital separation ($a$),
and eccentricity ($e$) distributions of binaries with the range of the parameter.
%$M_1$ denotes the primary mass and $q_1=M_1/M_2$ is the mass ratio where $M_2$ denotes the secondary mass.
For the star formation rate of Pop III stars, we use the same redshift dependence as that of \cite{DeSouza_2011},
but a value a factor of three smaller for the constraint of the Pop III star formation rate \citep{Inayoshi_2016}, which is  compatible with the cosmological Thomson scattering optical depth of the cosmic microwave background (CMB) measured by \cite{Planck_2016a}. 
To estimate the event rate of GW190521-like binaries, we use $10^6$ Pop III binaries for each model and pick up binary BHs with chirp masses that exceed $56\,\msun$, because the chirp mass of GW190521 is estimated as $56$--$77\,\msun$~\citep{Abbott:2020tfl}.

\begin{figure}
    \centering
    \includegraphics[width=0.47\textwidth]{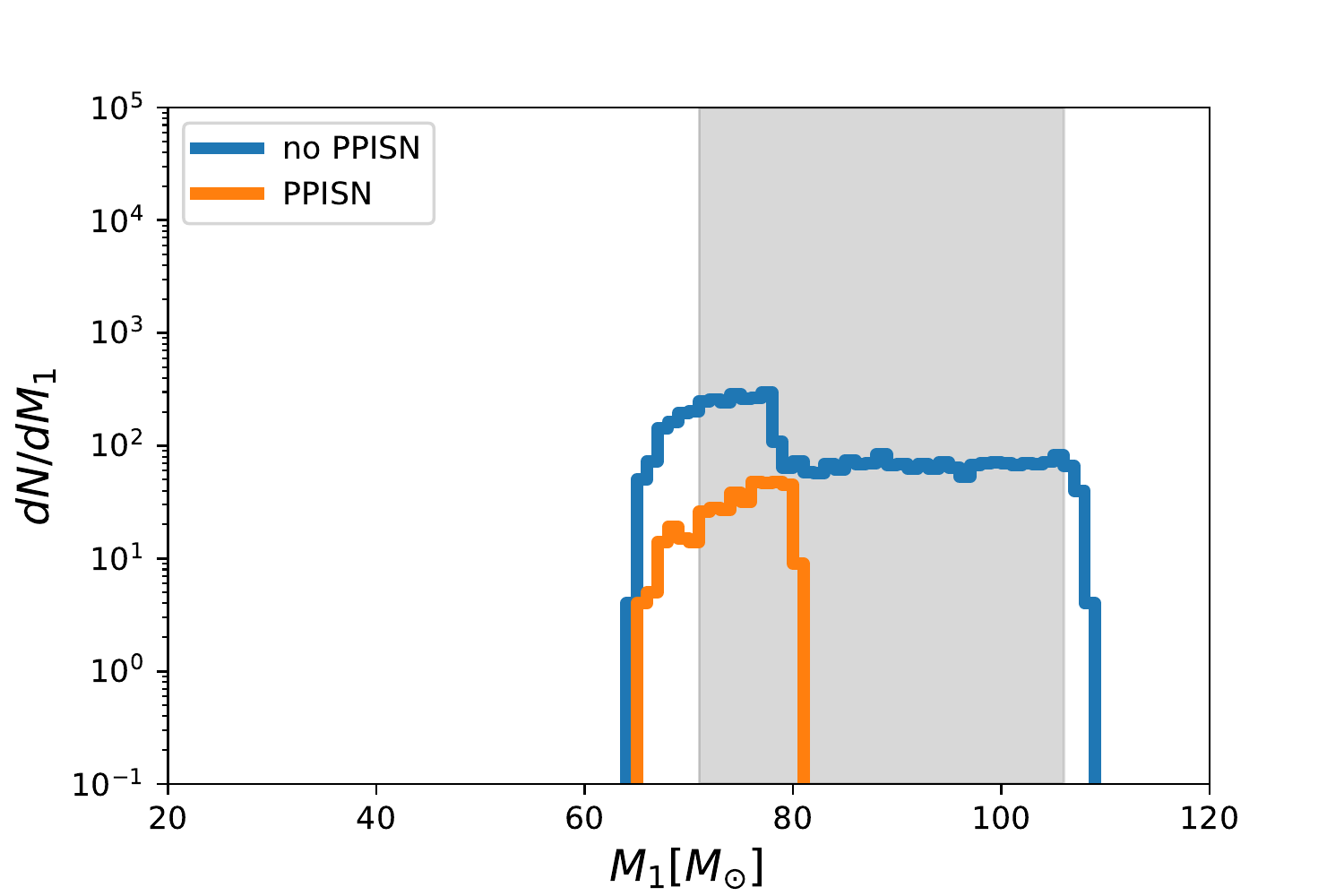}
    \caption{Primary BH mass distribution for binary BHs with chirp mass of $\geq 56\,\msun$
    that merge within the Hubble time.
    The blue line shows the model without mass ejection at PPISN (no PPISN), and the orange line shows the model with mass ejection (PPISN).
    The grey area is the mass range of the primary BH of GW190521.
    }
    \label{fig:primary}
\end{figure}

\begin{figure}
    \centering
    \includegraphics[width=0.47\textwidth]{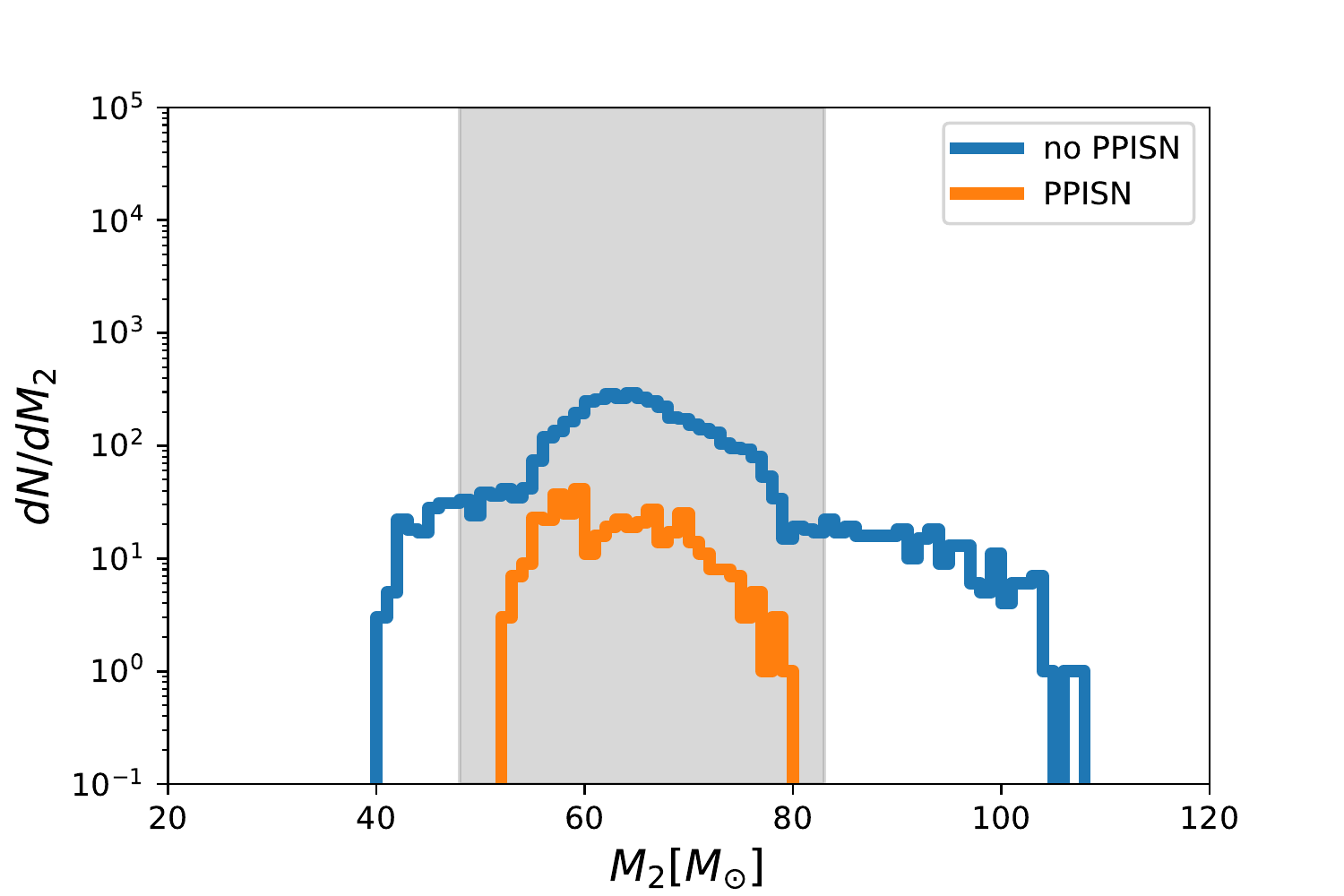}
    \caption{Same as Figure~\ref{fig:primary}, but for the secondary BH mass distribution and the mass range of the secondary BH of GW190521.
    }
    \label{fig:secondary}
\end{figure}

Figures~\ref{fig:primary} and \ref{fig:secondary} show the mass distribution of primary and secondary BHs with chirp masses that exceed 56$\,\msun$ and merge within the Hubble time.
If we consider perfect envelope loss due to PPISN, the maximum BH mass is $\sim80\,\msun$ from the orange lines
in Figures~\ref{fig:primary} and \ref{fig:secondary}.
Some Pop III stars with ZAMS masses of $\sim90$--$100\,\msun$, which avoid the PPISN and lose part of the envelope via mass transfer, can be the progenitors of such massive BHs.
On the other hand, if there is no envelope loss by Pop III PPISN, the maximum mass reaches $\sim105\,\msun$ from the blue lines in Figures~\ref{fig:primary} and \ref{fig:secondary}. Note that the grey areas in Figures~\ref{fig:primary} and \ref{fig:secondary} are the mass ranges of the primary and secondary BHs of GW190521, respectively. Thus, our two extreme theoretical predictions shown by the blue and orange lines are both consistent with the observed mass values of the primary and secondary BHs. 

{Here, let us define $\rho_{\rm SFR}$ and $Err_{\rm sys}$ by the cumulative comoving mass density of Pop III stars depending on the star formation rate, and the systematic error depending on uncertainties in the Pop III binary parameters. We first consider the case with $\rho_{\rm SFR}=(6\times10^5\msun/{\rm Mpc}^3)$ and $Err_{\rm sys}=1$.
In the no-mass ejection model, the current event rate of GW190521-like binary BH mergers is estimated as 
0.66$~{\rm yr^{-1}~Gpc^{-3}}$. In contrast, the perfect PPISN model gives 0.13$~{\rm yr^{-1}~Gpc^{-3}}$. Given that these two values originate from the extreme theoretical models, our fiducial model is defined as
a model between the two extreme models with a rate from 0.13 to 0.66$~{\rm yr^{-1}~Gpc^{-3}}$. The event rate of the general model can be expressed as 0.13--0.66$~(\rho_{\rm SFR}/(6\times10^5\msun/{\rm Mpc}^3)) Err_{\rm sys}~{\rm yr^{-1}~Gpc^{-3}}$. 
%Our rate is proportional to the Pop III SFR and $Err_{\rm sys}$.
$\rho_{\rm SFR}=6\times10^5\msun/\rm Mpc^3$ corresponds to the cumulative mass density of Pop III stars calculated by \cite{Inayoshi_2016}, assuming a flat IMF.
\cite{Kinugawa:2015nla} checked the dependence of $Err_{\rm sys}$ on binary parameters such as the IMF, common envelope parameter, mass loss fraction of the mass transfer, and BH natal kick.
They obtained the worst model in which the worst combination of parameters is used.
In the worst model, $Err_{\rm sys}\sim0.05$, while $Err_{\rm sys}\sim3.4$ in the best model. Using these estimates, we find that the event rate is in the range 0.0065--2.24$~(\rho_{\rm SFR}/(6\times10^5\msun/{\rm Mpc}^3))~{\rm yr^{-1}~Gpc^{-3}}$.
Except for near the worst model or the smallest-mass-density model, the theoretical rate is consistent with the observed range of 0.02--0.43$~{\rm yr^{-1}~Gpc^{-3}}$ from GW190521.}
%although the dependence on the SFR and $Err_{\rm sys}$.}

Finally, we estimate
the maximum observable redshift $z_{\rm max}$
for the LIGO O3a-Livingston (O3a-L),
LIGO O5,
Einstein Telescope (ET-B),
and Cosmic Explorer (CE2).
Using the inspiral--merger--ringdown waveform shown
in~\cite{Nakamura:2016hna},
we calculate the SNR
of GW events in fitted sensitivity curves for the
O3a-L ($f_{\rm low}=10\,$Hz),
O5 ($f_{\rm low}=10\,$Hz),
ET-B ($f_{\rm low}=1\,$Hz),
and CE2 ($f_{\rm low}=5\,$Hz)
used in~\cite{Kinugawa:2020tbg}.
Here, $f_{\rm low}$ is the lower frequency cutoff
and we set the higher frequency cutoff
at $f_{\rm high}=3000$\,Hz, although we do not need
such a high frequency for the heavy binary BHs considered below.
Subsequently, the maximum observable redshift
by setting the averaged SNR $=8$ 
is obtained for a binary BH with masses
($75\,M_{\odot}$, $75\,M_{\odot}$) as
$0.709$ for O3a-L,
$1.60$ for O5,
$10.8$ for ET-B,
and $19.3$ for CE2.
For a binary BH with masses ($80\,M_{\odot}$, $80\,M_{\odot}$),
$z_{\rm max}$ becomes
$0.734$ for O3a-L,
$1.61$ for O5,
$10.6$ for ET-B,
and $18.4$ for CE2.
The above calculations are summarized in Table~\ref{tab:range}.

\begin{table}
\caption{Maximum observable redshift $z_{\rm max}$
for GW190521-like binary BHs
for four ground-based GW detector configurations:
LIGO O3a-Livingston (O3a-L), LIGO O5 (O5),
Einstein Telescope (ET-B), and Cosmic Explore (CE2).
The mass is shown in solar masses $M_{\odot}$.}
\label{tab:range}
\begin{center}
\begin{tabular}{ccccc}
\hline
($M_1,\,M_2$) & O3a-L & O5 & ET-B & CE2 \\
\hline
(75,\,75) & 0.709 & 1.60 & 10.8 & 19.3 \\
(80,\,80) & 0.734 & 1.61 & 10.6 & 18.4 \\
\hline
\end{tabular}
\end{center}
\end{table}

%%%%%%%%%%%%%%%%%%%%%%%%%%%%%%%%%%%%%%%
\section{Discussion}
%%%%%%%%%%%%%%%%%%%%%%%%%%%%%%%%%%%%%%%

In the population synthesis simulations of Pop III
binary stars, we can simultaneously explain
the formation of 
binaries that consist of a BH and mass-gap compact object 
(MGCO) with a mass of 2--5$\,M_{\odot}$~\citep{Kinugawa:2020tbg}
like GW190814,
and those that consist of BHs with masses of $\sim 80\,M_{\odot}$
within the PISN mass gap for Pop III stars like GW190521. 
The first detected GW event, GW150914, the BH masses of which are $\sim30\,\msun$, had been predicted before
its first detection by \cite{Kinugawa2014}.
It is of great interest that our Pop III model can interpret the origins of three different classes of GW 
sources, that is, massive binary BHs with masses of $\sim30\,\msun$ like GW150914, BH and MGCO binaries with masses of $\sim 2.5\,\msun$ like GW190814,
and very massive binary BHs with masses of $\sim 80\,\msun$ like GW190521.
It will be triple the fun! 
{If the origin of massive binary BHs is Pop III field binaries, the merger rate density has a peak at $z\sim10$ \citep{Kinugawa:2020tbg}.
Thus, the future work of GW observatories such as ET \citep{ET}, CE \citep{Reitze:2019iox}, and DECIGO \citep{Seto:2001qf} can verify whether our model is correct or not.}

The Pop III binary BH merger rate strongly depends on the Pop III star formation rate (SFR) and IMF, which are still observationally uncertain because no Pop III star has ever been observed. 
In the case of the Pop III SFR, there is a constraint on the density of gas from the Thomson scattering optical depth  obtained from CMB observation \citep{Visbal_2015, Inayoshi_2016}.
%A comparison between the observation rate and the theoretical rate might give the constraint of Pop III SFR.
On the other hand, the merger rate also depends on the Pop III IMF.
In our previous study \citep{Kinugawa:2015nla}, we verified the IMF dependence of Pop III binary BHs using the flat, logflat, and Salpeter IMFs.
The number of merging massive binary BHs ($M_{\rm total}\sim150\,\msun$) for the logflat IMF decreases to a value about two thirds that of the flat IMF.
Although we have no observational data of Pop III stars, some numerical simulations and semi-analytical calculations have been performed to estimate the Pop III IMF.
{\cite{Hirano_2014} and \cite{Susa_2014} showed the Pop III stellar formation in star-forming minihalos using cosmological simulations, and obtained the Pop III IMF for $\sim10$--$1000\msun$.
Recent simulations \citep{Hirano2017,Susa2019} suggest that fragmentation causes the typical mass of Pop III stars to be small, although there are still some uncertainties such as the merging of fragments and the magnetic effect.}

{On the other hand, \cite{Tarumi2020} have estimated the Pop III IMF from the metallicity distribution analysis of the inhomogeneous metal mixing effect calibrated by the metallicity distribution function of extremely metal-poor (EMP) stars.
They showed that the Pop III IMF is proportional to $M^{-0.5}$~($2$--$180\msun$), which is different from the flat IMF adopted in this study. To compare the effect of different IMFs, in this study we employ a simple method using the probability, although population synthesis calculations are needed for detailed analysis. As shown in the footnote, the probability of having GW190521-like binary BHs for the Tarumi IMF is approximately $78\%$ of that for our flat
IMF~\footnote{{Let us choose the minimum and maximum masses of Pop III stars as $M_{\rm min}=10 M_{\odot}$ and $M_{\rm max}=150 M_{\odot}$, respectively. We also define IMFs for our case and the Tarumi case as $f_{\rm flat}(M)=C_{\rm flat}=$~constant and $f_{\rm Tarumi}(M)=C_{\rm Tarumi}/M^{0.5}$, respectively. By fixing the total number of stars ($N_{\rm tot}$), we have $C_{\rm flat}=N_{\rm tot}/(M_{\rm max}-M_{\rm min})$ and $C_{\rm Tarumi}=N_{\rm tot}/[2\{(M_{\rm max})^{0.5}-(M_{\rm min})^{0.5}\}]$. The probability of a star with a mass $M$ in the interval $dM$ is proportional to $f(M)\,dM$, where $f(M)$ is an IMF. Therefore, the relative probability ($p(M)$) of a star of mass $M$ that exists in the Tarumi IMF compared with the flat IMF is given by
$p(M)=f_{\rm Tarumi}(M)/f_{\rm flat}(M)$.
Using $p(M)$, the relative probability of the existence of a binary with masses
$M_1$ and $M_2$ is equal to $p_{\rm b}(M_1,M_2)=p(M_1)p(M_2)$ if there is no correlation. For GW190521, $M_1\sim 88M_{\odot}$ and $M_2\sim 65M_{\odot}$, and thus we obtain $p_{\rm b}(M_1,M_2)\sim 0.78$.}}.
However, this is a simple order-of-magnitude argument and more population synthesis simulations are needed for a wider class of possible IMFs. Theoretically, when we consider different IMFs, it is not clear which macroscopic quantity is fixed: the total number of stars, the total mass, or the total luminosity. This problem should also be solved in the future. 
The future work of GW observatories such as ET, CE, and DECIGO can detect massive binary BH mergers at 
$z\gtrsim10$. The comparison between the binary BH mass distribution at high redshift and theoretical estimations of the Pop III IMF might reveal the Pop III stellar mass distribution.}

\cite{Liu:2020krj} have recently considered another possibility of Pop III binary BH mergers originating from dynamical capture. They calculated the merger rate of Pop III binary BHs made by dynamical capture in cosmological hydrodynamic simulations. Although their merger rate ($0.04~{\rm yr^{-1}~Gpc^{-3}}$) is smaller than that for our field binary case, {the Pop III binary BHs formed by dynamical capture might have different features from Pop III binary BHs made by field binaries, such as large eccentricity by dynamical friction and larger typical mass of BHs, because of the absence of mass loss in binary interactions.}
These differences might be observed by ET or other future GW observations.
The origin of binary BHs will be made clear by the GW observation of
the redshift $z>10$ (see Table~\ref{tab:range}),
i.e., in the ET-B era.
This is because the cumulative event rate of Pop I/II binary BHs saturates
at $z<5$, and that of Pop III binary BHs saturates at $z \sim 10$~\citep{Nakamura:2016hna}.

In this paper, we did not focus on the spin values
of binary BHs.
The spin estimation for GW190521 has a large uncertainty
as the nondimensional spin parameters $\chi_1=$~0.07--0.96
and $\chi_2=$~0.09--0.97, although there is a weak preference
for a spinning, precessing binary BH, i.e., the BH spins
may be misaligned from the orbital angular momentum.
Furthermore, according to \cite{P2000158-v4},
the orientation of spins projected on the orbital plane
is not determined.
By observing the long inspiral phase
with space-based GW detectors
(see, e.g., B-DECIGO in~\cite{Isoyama:2018rjb}
and TianQin in~\cite{Mei:2020lrl},
we will be able to access more precise information
on the spins.
{Whether Pop III stars rotate or not is related to the abundance patterns of EMP stars \citep{Takahashi2014, Choplin2019}. 
If we can obtain more precise information on the spins from the future work of space-based GW detectors,
it might be possible to impose a constraint on the Pop III stellar rotation, and to consider the rotation effect on the abundance patterns of EMP stars.}

%%%%%%%%%%%%%%%%%%%%%%%%%%%%%%%%%%%%%%%
\section*{Acknowledgment}
%%%%%%%%%%%%%%%%%%%%%%%%%%%%%%%%%%%%%%%
{We would like to gleatly thank the anonymous referee for their useful comments to improve our paper. } 
We thank Ataru Tanikawa and Takahiro S. Yamamoto for useful discussions on the mass-gap BH and the BH spins of GW190521.
T. K. acknowledges support from the University of Tokyo Young Excellent Researcher program.
T. N. acknowledges support from
JSPS KAKENHI Grant No. JP15H02087.
H. N. acknowledges support from
JSPS KAKENHI Grant Nos. JP16K05347 and JP17H06358.
{We thank Editage (www.editage.com) for English language editing.}
\section*{Data Availability}
Results will be shared on reasonable request to corresponding author.
%%%%%%%%%%%%%%%%%%%%%%%%%%%%%%%%%%%%%%%%
\bibliographystyle{mnras}

\bibliography{ref}

%%%%%%%%%%%%%%%%%%%%%%%%%%%%%%%%%%%%%%%%
\end{document}